\documentstyle[12pt]{article}
\begin{document}
\begin{titlepage}
\title{\bf{Transformations of real-time\\
finite-temperature Feynman rules}}
\author{M.~A.~van Eijck\\
Institute for Theoretical Physics,
University of Amsterdam,\\
Valckenierstraat 65,
NL-1018 XE Amsterdam,
The Netherlands\vspace*{8pt}\\
R.~Kobes\\
Laboratoire de Physique Th\'eorique ENSLAPP,
B.~P.~110\\  F--74941 Annecy-le-Vieux Cedex, France\\
and\\
Physics Department and Winnipeg Institute
for Theoretical Physics,\\
University of Winnipeg,
Winnipeg, Manitoba R3B 2E9, Canada\vspace*{8pt}\\
Ch.~G.~van Weert\\
Institute for Theoretical Physics,
University of Amsterdam,\\
Valckenierstraat 65,
NL-1018 XE Amsterdam,
The Netherlands}
\maketitle
\vskip 0.5truein
\begin{abstract}
We consider transformations of the $2\times2$ propagator
matrix in real-time finite-temperature field theory, resulting in
transformed $n$--point functions.
As special cases of such a transformation we examine
the Keldysh basis, the retarded/advanced $RA$ basis, and a Feynman-like
$F\bar F$ basis, which differ in this context as to
how ``economically'' certain constraints on the original
propagator matrix elements are implemented. We also obtain the
relation between some of these real-time functions and certain analytic
continuations of the imaginary-time functions. Finally, we compare
some aspects of these bases which arise
in practical calculations.
\end{abstract}
\begin{flushright}
ITFA--94--15\\
ENSLAPP--A--471/94\\
NSF--ITP--94--61\\
\end{flushright}
\thispagestyle{empty}
\end{titlepage}

\section{Introduction}
The fields $\phi(x)=\phi(t,\vec{x})$ describing particles in a heat
bath at temperature $\beta^{-1}$ are either periodic (bosons)
or anti-periodic (fermions) in the complex time plane,
with period $-i\beta$ [1--5].
Integrals over time then follow a
path $C$ in the complex time plane connecting a point
$-T$ to a point $-T-i\beta$. In the imaginary-time formalism
the straight path connecting these points is chosen, resulting in
time arguments $t=-i\tau$ that are purely imaginary -- an analytic
continuation is then involved for quantities defined for real times.

In real-time formalisms a path
which includes the real time axis is chosen,
and thermal Green functions are then
time-ordered with respect to this path:
\begin{equation}
G_C(x_1,\dots,x_n)=\left<T_C\phi(x_1)\cdots\phi(x_n)\right>.
\end{equation}
A standard choice of this path is the concatenation of the
following four pieces:
$C_I:\ -T\to T;~~C_{III}:\ T\to T-i\sigma;~~
C_{II}:\ T-i\sigma\to -T-i\sigma;~~C_{IV}:\ -T-i\sigma\to -T-i\beta$,
where $\sigma$ is an arbitrary parameter $0\leq\sigma\leq\beta$
\cite{land,um}.
As $T\rightarrow\infty$ the contributions
from $C_{III}$ and $C_{IV}$ can be neglected in the majority of cases.
It is then convenient to introduce the thermal doublet
\begin{equation}
\phi_a=\left(\begin{array}{c}\phi_+\\\phi_-\end{array}\right),
\end{equation}
where $a,b=+,-$.
The fields $\phi_+(x)=\phi(x)$ and $\phi_-(x)=\phi(t-i\sigma,\vec{x})$
are defined on the forward and backward segments of the time contour,
$C_I$ and $C_{II}$ respectively.
This leads to a matrix form of the thermal Green function:
\begin{equation}
G_{a_1\cdots a_n}(x_1,\dots,x_n)=
\left<T_C\phi_{a_1}(x_1)\cdots\phi_{a_n}(x_n)\right>.
\end{equation}
The dependence on $\sigma$ will not be indicated explicitly,
except in cases where a particular value applies.
Two popular choices for this contour parameter that have been studied are
$\sigma=0$, which is the closed time-path formalism of
Schwinger, Keldysh and others [6--9], and $\sigma=\frac{1}{2}$, which
is equivalent to the ``unitary'' formulation
of thermo field dynamics \cite{um}.

The propagator for free particles is a $2\times2$ matrix and in momentum
space it has the form
\begin{equation}
D_{ab}(p)=\left( \begin{array}{cc}
D_{++}(p) & D_{+-}(p)\\
D_{-+}(p) & D_{--}(p) \end{array}\right),
\label{delta}\end{equation}
where
\begin{eqnarray}
& & D_{++}(p)=D_{--}^*(p)
= \nonumber\\& &\qquad=\theta(p_0)\Delta_F(p)-\theta(-p_0)\Delta_F^*(p)
+\eta\varepsilon(p_0)n(x)[\Delta_F(p)+\Delta_F^*(p) ], \nonumber\\
& & D_{+-}(p)=\eta e^{2\sigma p_0-x}D_{-+}(p)
= \nonumber\\& &\qquad=\eta\varepsilon(p_0)n(x)e^{\sigma p_0}
[\Delta_F(p)+\Delta_F^*(p) ].
\label{explicitDelta}\end{eqnarray}
Here $\Delta_F(p)$ is the Feynman propagator
($i/[p^2-m^2+i\varepsilon]$ for scalar bosons), and
\begin{equation}
n(x)=\frac{1}{e^x-\eta}
\end{equation}
is the Bose-Einstein (Fermi-Dirac) distribution function with the properties
\begin{eqnarray}
& &e^x n(x)=-\eta n(-x),\nonumber\\
& & n(x)+n(-x)=-\eta,
\end{eqnarray}
where $x=\beta(p_0-\mu)$, $\mu$ is the chemical potential, and
$\eta=+1\,(-1)$ for bosons (fermions).

In many applications Green functions other than these time-ordered
products arise. A common example is the linear response of the field
to a weak external current \cite{fw,chou},
which involves the retarded propagator $\Delta_R$.
This product also arises in a comparison of certain analytic
continuations of the $n$--point functions in the imaginary-time
formalism to functions in real-time formalisms
\cite{land,evans1,evans2,nieg,guer2}.
As the retarded propagator can be written as
\begin{equation}
\Delta_R(p)=
D_{++}(p)-e^{-\sigma p_0}D_{+-}(p)
=\theta(p_0)\Delta_F(p)-\theta(-p_0)\Delta_F^*(p),
\label{retdef}\end{equation}
which for bosons is $i/[p^2-m^2+i\varepsilon p_0]$, it is then of
interest to consider transformations of Eq.~(\ref{delta}) which
could generate such linear combinations of the $D_{ab}$.

In this paper we will examine transformations
to a ${\widehat D}_{XY}$ basis of the form
\begin{equation}
 U(p)D(p)U^T(-p)={\widehat D}(p),
\label{mat} \end{equation}
where ``$T$'' denotes the transpose and
\begin{equation}
U_{Xa}(p)=\left( \begin{array}{cc}
f_{1+}(p) & f_{1-}(p)\\
f_{2+}(p) & f_{2-}(p)  \end{array}\right).
\end{equation}
Without further input, however, there would likely be no advantage
to study such a transformation in general.
One observation that could be a guide for choosing
``useful'' transformations is that since the original propagators
$D_{ab}(p)$ satisfy the relations \cite{land}
\begin{eqnarray}
& & D_{++}(p)+D_{--}(p)-e^{-\sigma p_0}D_{+-}(p)
-e^{\sigma p_0}D_{-+}(p)=0, \label{rel1} \\
& & D_{++}(p)+D_{--}(p)-\eta e^{\sigma p_0-x}D_{-+}(p)
-\eta e^{x-\sigma p_0}D_{+-}(p)=0, \label{rel2}
\end{eqnarray}
the transformed
${\widehat D}_{XY}(p)$ of Eq.~(\ref{mat}) will also
not be independent in general.
It could then be of some advantage to
choose the transformation so that either one or both of
the relations of Eqs.~(\ref{rel1}, \ref{rel2})
translate into the trivial vanishing of
either one or two of the ${\widehat D}_{XY}(p)$.

It may be noted that the transformation of Eq.~(\ref{mat}) has a close
correspondence to the Bogoliubov transformation
studied in thermo field dynamics \cite{henn}.
In fact, the transformation matrices $U(p)$ can be mapped onto a
particular set of Bogoliubov transformations seeking
a diagonal form for the propagator by a transformation of
the field operators \cite{cvw}.

\section{Transformed $n$--point functions}
The first relation of Eq.~(\ref{rel1}) among the $D_{ab}(p)$ elements
is an immediate consequence of the definition of the
path-ordered product, whereas for the second one of Eq.~(\ref{rel2})
one also needs the Kubo-Martin-Schwinger (KMS) condition for
equilibrium two--point correlation functions. These relations are
thus also satisfied by the full propagator $G_{ab}(p)$. Hence,
the full propagator has the same matrix components of Eq.~(\ref{delta})
as the bare one,
with now the Feynman propagators in Eq.~(\ref{explicitDelta}) given by
the spectral representation
\begin{equation}
\Delta_F(p)=\int_0^\infty d\tau^2
\frac{\rho(\tau,\vec{p}\,)}{p_0^2-\tau^2+i\varepsilon}.
\end{equation}
This implies that for the full propagator we may write the spectral form
\begin{equation}
G_{ab}(p)=\int_0^\infty d\tau^2\rho(\tau,\vec{p}\,)D_{ab}(\tau,\vec{p}\,)
\end{equation}
and the analogous one for the transformed propagator
\begin{equation}
{\widehat G}_{XY}(p)=\int_0^\infty d\tau^2\rho(\tau,\vec{p}\,)
{\widehat D}_{XY}(\tau,\vec{p}\,),
\end{equation}
owing to the identical matrix structures of interacting and
non-interacting propagators.

Consider now Dyson's equation for the full
propagator defined on the contour $C_I\cup C_{II}$ \cite{land,chou}:
\begin{equation}
\int_C dy\, \Gamma_C(x,y)G_C(y,z)=\delta_C(x-z),
\end{equation}
or, in a matrix representation,
\begin{equation}
\int_{-\infty}^{+\infty}
 dy\, \Gamma(x,y)\sigma_3 G(y,z)=\sigma_3\delta(x-z),
\end{equation}
where $\sigma_3=$ diag$(1,-1)$ is the $2\times2$ Pauli matrix.
Writing $\Gamma_C(x,y)=\Gamma^{(0)}_C(x,y)-\Sigma_C(x,y)$,
in momentum space we then have
\begin{equation}
G(p)=D(p)+D(p)\Sigma(p)G(p),
\label{dyson}\end{equation}
where here, and also in the following for $n$--point functions,
the effects of $\sigma_3$ have been absorbed
into the vertex function by associating a minus sign with
each ``$-$'' index.
 From Eq.~(\ref{mat}) and the preceding discussion we thus see
\begin{equation}
U(p)G(p)U^T(-p)={\widehat G}(p),
\end{equation}
where ${\widehat G}(p)$ satisfies
\begin{equation}
{\widehat G}(p)={\widehat D}(p)+
{\widehat D}(p){\widehat\Sigma}(p){\widehat G}(p)
\end{equation}
and the transformed
(amputated) self-energy ${\widehat\Sigma}(p)$ is given by
\begin{equation}
V(p)\Sigma(p)V^T(-p)={\widehat\Sigma}(p),
\label{setran}\end{equation}
where we have defined
\begin{equation}
V_{Xa}(p)=\left( \begin{array}{cc}
V_{1+}(p) & V_{1-}(p)\\
V_{2+}(p) & V_{2-}(p) \end{array}\right)
\equiv\left[U^T(-p)\right]^{-1}.
\label{v}\end{equation}

Generalizing this we can, as with Aurenche and Becherrawy \cite{aur1,mich},
introduce transformed (amputated) $n$--point
functions ${\widehat\Gamma}_{X_1\cdots X_n}(p_1,\ldots,p_n)$
in terms of the $+/-$ functions $\Gamma_{a_1\cdots a_n}
(p_1,\ldots,p_n)$. If all momenta are incoming, such functions are
\begin{equation}
{\widehat\Gamma}_{X_1\cdots \cdots X_n}
(p_1,\ldots,p_n)=V_{X_1 a_1}(p_1)
\cdots V_{X_n a_n}(p_n)
\Gamma_{a_1\cdots a_n}(p_1,\ldots,p_n).
\label{tran}\end{equation}
If some of the momenta were outgoing, say $p_{k+1},\ldots,p_n$,
the matrix $V^T(-p)$ rather than $V(p)$
would be associated with these momenta in Eq.~(\ref{tran}).
Because of this, the corresponding
transformed $n$--point function is
then simply related to that with all momenta incoming by
\begin{equation}
{\widehat\Gamma}_{X_1\cdots X_k;X_{k+1}\cdots X_n}
(p_1,\ldots,p_k;p_{k+1},\ldots,p_n)=
{\widehat\Gamma}_{X_1\cdots X_n}
(p_1,\ldots,p_k,-p_{k+1},\ldots,-p_n).
\end{equation}
Thus, it suffices to study these functions with all
momenta incoming. We will assume this in the following,
so that $\sum p_{i_0}=0=\sum \mu_i$, and we also use fermion number
conservation: $\prod \eta_i =1$. Also in what follows
the delta functions associated with these conservation laws have been
factored out from the Green functions.

Feynman rules for the transformed functions of Eq.~(\ref{tran})
then follow from the rules for the $n$--point functions
$\Gamma_{a_1\cdots a_n}$, which involve internal propagators $D_{ab}$
connected by appropriate bare vertices $g_{ab\cdots}$:
\begin{equation}
g_{+\cdots +}(p_1,\ldots,p_n)=-g_{-\cdots -}(p_1,\ldots,p_n)\equiv
g(p_1,\ldots,p_n),
\label{vertices}\end{equation}
with all others being zero \cite{land}.
Instead of these vertices one may also consider amputated
subdiagrams, for which the following discussion will also apply.
The internal propagators in a graph could be written in
terms of transformed propagators ${\widehat D}_{XY}$ using
Eq.~(\ref{mat}) and, as
also done by Aurenche and Becherrawy \cite{aur1},
the matrices of each such transformation could be associated
with the adjacent vertices. This will ultimately
replace each vertex
$g_{ab\cdots}$ in a graph by the appropriate
transformed vertex ${\widehat g}_{XY\cdots}$,
where the transformation is given in Eq.~(\ref{tran}).
The end result is that the Feynman rules
for the transformed $n$--point functions
can be written solely in terms of
transformed vertices connected by transformed internal
propagators.

A simple illustration of this transformation of
Feynman rules is a one-loop self-energy graph
involving cubic interactions such as
\begin{equation}
-i\Sigma_{aa'}(p)=(-i)^2\int\, {d^4k\over (2\pi)^4}\
g_{abc}(p,-k,q)D_{bb'}(k)g_{a'b'c'}(-p,k,-q)
D_{c'c}(q),
\end{equation}
where $q=k-p$ and only thermal indices are retained.
The transformed self-energy ${\widehat\Sigma}(p)$ of
Eq.~(\ref{setran}) corresponding to this graph is, using Eq.~(\ref{mat}),
\begin{eqnarray}
& &-i {\widehat\Sigma}_{XX'}(p)
=V_{Xa}(p)\left[ -i\Sigma_{aa'}(p) \right]
V_{X'a'}(-p)=\nonumber\\
& &\quad=(-i)^2\int\, {d^4k\over (2\pi)^4}\
{\widehat g}_{XYZ}(p,-k,q)
{\widehat D}_{YY'}(k){\widehat g}_{X'Y'Z'}(-p,k,-q)
{\widehat D}_{Z'Z}(q),
\label{selfenergy}\end{eqnarray}
where, as in Eq.~(\ref{tran}),
\begin{eqnarray}
& &{\widehat g}_{XYZ}(p,-k,q)=
V_{Xa}(p) V_{Yb}(-k) V_{Zc}(q) g_{abc}(p,-k,q) ,\nonumber\\
& &{\widehat g}_{X'Y'Z'}(-p,k,-q)=
V_{X'a'}(-p) V_{Y'b'}(k) V_{Z'c'}(-q)  g_{a'b'c'}(-p,k,-q).
\end{eqnarray}

It will be found convenient in the construction of the following
transformations to aim for the $+/-$ $n$--point functions
$\Gamma_{a_1\cdots a_n}$ to be multiplied by one of two
particular factors for each index $a_i$ with the value ``$-$'':
\begin{eqnarray}
& &\Gamma_{a_1\cdots a_n}(p_1,\ldots,p_n)\prod_{a_i=-}
e^{\sigma p_{i_0}}, \nonumber \\
& & \Gamma_{a_1\cdots a_n}(p_1,\ldots,p_n)\prod_{a_i=-}
\eta_i e^{\sigma p_{i_0}-x_i}.
\end{eqnarray}
The exponential factors exactly eliminate the
$\sigma$--dependence of $\Gamma_{a_1\cdots a_n}$, leaving
these combinations $\sigma$--independent.
The first function here is simply the $\sigma=0$
$n$--point function expressed in terms of $\Gamma_{a_1\cdots a_n}$
with an arbitrary $\sigma$.
The second function can be related to the first
by the KMS relation.
These functions satisfy
\begin{eqnarray}
& &\sum_{a_i=\pm}\left[
\Gamma_{a_1\cdots a_n}(p_1,\ldots,p_n)\prod_{a_i=-}
e^{\sigma p_{i_0}}\right]=0,\nonumber\\
& &\sum_{a_i=\pm}\left[
\Gamma_{a_1\cdots a_n}(p_1,\ldots,p_n)\prod_{a_i=-}
\eta_i e^{\sigma p_{i_0}-x_i}\right]=0,
\label{lte}\end{eqnarray}
which can be proven, for example, using the largest and smallest
time equations of 't Hooft and Veltman \cite{vel,circ}.
These relations are the amputated versions of the $n$--point generalizations
of the 2--point relations of Eqs.~(\ref{rel1}, \ref{rel2}).
Note that, unlike Eqs.~(\ref{rel1}, \ref{rel2}),
in Eq.~(\ref{lte}) no minus signs appear due to the
their absorption into the amputated vertex functions,
as discussed after Eq.~(\ref{dyson}).

We shall also have occasion to consider the particular combinations
\begin{eqnarray}
& & \Gamma^{(R)}(p_1;p_2,\ldots,p_n)\equiv
\sum_{a_i=\pm}\left[ \Gamma_{+a_2\cdots a_n}(p_1,\ldots,p_n)
\prod_{a_i=-}
e^{\sigma p_{i_0}}\right],\nonumber\\
& & \Gamma^{(A)}(p_1;p_2,\ldots,p_n)\equiv
\sum_{a_i=\pm}\left[ \Gamma_{+a_2\cdots a_n}(p_1,\ldots,p_n)
\prod_{a_i=-}
\eta_i e^{\sigma p_{i_0}-x_i}\right],
\label{retadv}\end{eqnarray}
which are real-time representations of the one-particle-irreducible
contributions to
the generalized $n$--point retarded and advanced
products \cite{chou,mich,resp}.
They are related by complex conjugation: $\Gamma^{(A)}(p_1;p_2,\ldots,p_n)
=\Gamma^{(R)\,\,*}(p_1;p_2,,\ldots,p_n)$,
which can be shown using Eq.~(\ref{lte})
together with the conjugation relation \cite{mich}
\begin{equation}
\Gamma_{a_1\cdots a_n}^*(p_1,\ldots,p_n)=
-\Gamma_{{\bar a_1}\cdots {\bar a_n}}(p_1,\ldots,p_n)
\prod_{a_{i}=-} \eta_i e^{x_i-2\sigma p_{i_0}},
\label{pmcc}\end{equation}
where ${\bar a}=-,\,+$ when $a=+,\,-$. As will be seen
later, the identification of $\Gamma^{(R)}$ and
$\Gamma^{(A)}$ with the retarded and advanced amputated
functions, respectively, is based
on the convention in the imaginary-time formalism
for the retarded and advanced analytic continuations of
the (non-amputated) Green functions \cite{evans1,evans2,nieg,guer2}.

\section{The Keldysh basis}
The Keldysh transformation of the time-path method is an example where
one of the relations of Eqs.~(\ref{rel1}, \ref{rel2}) is used
to eliminate one of the transformed functions
${\widehat D}_{XY}(p)$ [6--9].
Such a transformation is particularly useful in non-equilibrium
situations, where a relation such as Eq.~(\ref{rel2}), which follows from
the KMS condition, does not hold.
We will find that this class of transformations automatically involves
the retarded and advanced propagators
$\Delta_R(p)=-\Delta_A^*(p)$ of Eq.~(\ref{retdef}), but
there is some freedom in the form of
the third non--zero element. As is conventional, we
choose this element to be proportional to the
combination $D_{++}(p)+D_{--}(p)$.

We consider the case where the transformed
function ${\widehat D}_{22}(p)$ of Eq.~(\ref{mat}) vanishes
identically by Eq.~(\ref{rel1}).
With $k_i= 1,\,2$, the transformed propagators,
\begin{equation}
 {\widehat D}_{k_1 k_2}(p)=
\left( \begin{array}{cc}
 {\widehat D}_{11}(p) & {\widehat D}_{12}(p)\\
 {\widehat D}_{21}(p) &  {\widehat D}_{22}(p)
\end{array}\right)=
\left( \begin{array}{cc}
\Delta_{S}(p) & \Delta_R(p)\\
\Delta_A(p) & 0
\end{array}\right),
\label{kel}\end{equation}
with the normalization on the right--hand--side,
follow by choosing in Eq.~(\ref{mat})
\begin{equation}
U_{ka}(p)=\frac{1}{\sqrt{2}b(-p)}\left( \begin{array}{cc}
b(p)b(-p) & e^{\sigma p_0}b(p)b(-p)   \\
1 & -e^{\sigma p_0} \end{array}\right),
\label{uk}\end{equation}
where $b(p)$ is an arbitrary function and the
symmetric product $\Delta_{S}(p)$ is
\begin{equation}
\Delta_{S}(p)=b(p)b(-p)\left[D_{++}(p)
+D_{--}(p)\right] =  b(p)b(-p)
\coth^\eta(x/2)\left[\Delta_R(p)-
\Delta_A(p)\right].
\label{cor}\end{equation}

We then define ``1/2'' $n$--point functions
${\cal K}_{k_1 \cdots k_n}(p_1,\ldots,p_n)$
in terms of the $+/-$ $n$--point functions
$\Gamma_{a_1 \cdots a_n}(p_1,\ldots,p_n)$ by Eq.~(\ref{tran}).
The corresponding matrix $V(p)$ of Eq.~(\ref{v}) in this case is,
from Eq.~(\ref{uk}),
\begin{equation}
\left(\begin{array}{cc}
V_{1+}(p) & V_{1-}(p)\\
V_{2+}(p) & V_{2-}(p) \end{array}\right)=
\frac{1}{\sqrt{2}b(-p)}\left( \begin{array}{cc}
1 & e^{\sigma p_0}   \\
b(p)b(-p) &
-e^{\sigma p_0}b(p)b(-p) \end{array}\right).
\label{kelv}\end{equation}
In this notation the amputated 2--point functions are
\begin{eqnarray}
& & {\cal K}_{11}(p_1,p_2)=0,\nonumber\\
& & {\cal K}_{12}(p_1,p_2)=\Gamma^{(A)}(p_1;p_2),\nonumber\\
& & {\cal K}_{21}(p_1,p_2)=\Gamma^{(R)}(p_1;p_2),\nonumber\\
& & {\cal K}_{22}(p_1,p_2)=  b(p_1)b(p_2)
\coth^\eta(x_{p_1}/2)\left[\Gamma^{(R)}(p_1;p_2)-
\Gamma^{(A)}(p_1;p_2)\right],
\label{con}\end{eqnarray}
and in general one can show, using the largest/smallest time
equation relations of Eq.~(\ref{lte}), that
\begin{eqnarray}
& & {\cal K}_{11\cdots1}(p_1,\ldots,p_n)=0, \nonumber\\
& & {\cal K}_{211\cdots1}(p_1,\ldots,p_n)=
2^{1-n/2}\frac{b(p_1)}{b(-p_2)\cdots b(-p_n)}
\Gamma^{(R)}(p_1;p_2,\ldots,p_n),
\label{gen}\end{eqnarray}
where $\Gamma^{(R)}(p_1;p_2,\ldots,p_n)$
is defined in Eq.~(\ref{retadv}).

Feynman rules for the 1/2 functions consist of internal
propagators of Eq.~(\ref{kel}) and bare vertices
obtained from Eqs.~(\ref{tran}, \ref{kelv}) using the
bare $+/-$ vertices of Eq.~(\ref{vertices}):
\begin{equation}
\frac{ {\widehat g}_{k_1\cdots k_n}(p_1,\ldots,p_n)}
{g(p_1,\ldots,p_n)}=
\frac{1-(-1)^{\#2}}{2^{n/2} }
\prod_{k_i=1} \frac{1}{b(-p_i)}
\prod_{k_i=2} b(p_i),
\end{equation}
where ``$\#2$'' denotes the number of ``2'' indices present.
Thus, bare vertices with an even number of ``2'' indices vanish.
The non-vanishing ones satisfy the following relation when the signs of
all energies and momenta are reversed:
\begin{equation}
\frac{ {\widehat g}_{k_1\cdots k_n}(-p_1,\ldots,-p_n)}
{{\widehat g}_{k_1\cdots k_n}(p_1,\ldots,p_n)}=
\frac{ g(-p_1,\ldots,-p_n)}{g(p_1,\ldots,p_n)}
\prod_{k_i=1,2} \frac{b(-p_i)}{b(p_i)}.
\end{equation}
In these considerations there is no obvious advantage
to choosing the free parameter $b(p)$ to be momentum dependent,
and so choices such as $b(p)=1$ or $1/\sqrt{2}$ are usually
most convenient.

As in this type of transformation only the one constraint
of Eq.~(\ref{rel1}) satisfied by the $D_{ab}$
was used to eliminate one of the transformed Green functions,
the other constraint of Eq.~(\ref{rel2}) will lead to non-trivial
relations amongst the transformed functions ${\cal K}_{k_1\cdots k_n}$.
This is illustrated by the last
relation of Eq.~(\ref{con}) relating ${\cal K}_{22}$ to
${\cal K}_{12}$ and ${\cal K}_{21}$, which is referred to as the
fluctuation-dissipation theorem \cite{chou}.

One could also consider the case where the transformed
function ${\widehat D}_{22}(p)$ of Eq.~(\ref{mat}) vanishes
by Eq.~(\ref{rel2}), rather than by Eq.~(\ref{rel1}) as chosen here.
The analysis of such a transformation is straightforward;
we only note that in this case the transformed
$n$--point function ${\cal K}_{211\cdots 1}$ will be proportional
to $\Gamma^{(A)}$, rather than to
$\Gamma^{(R)}$ as happened in Eq.~(\ref{gen}).

These Keldysh-like transformations are convenient in a
general analysis of response theory \cite{chou}. In this approach one adds
to the Hamiltonian a term involving a (weak) external current $J(t)$
and then calculates the response of various Green functions.
Translating from the notation of Ref.~\cite{chou} into
that used here, one has
\begin{eqnarray}
& & \delta G_1(t)\sim \int\,dt_1\ G_{12}(t,t_1)J(t_1)
+\frac{1}{2} \int\,dt_1\,dt_2\ G_{122}(t,t_1,t_2)J(t_1)J(t_2)+\ldots,
\nonumber\\
& & \delta G_{11}(t,t')\sim \int\,dt_1\ G_{112}(t,t',t_1)J(t_1)+\ldots,
\label{resp}\end{eqnarray}
and so on. The Green functions $G_{122\cdots 2}(t_1,\ldots,t_n)$
with one ``1'' index
are the $n$--point retarded products,
and arise in the response of the
expectation value $G_1(t)$ of the field operator, while
those response functions with more than one ``1'' index
arise in the response of higher-order correlation functions
$G_{11\cdots1}$. The one-particle
irreducible functions ${\cal K}_{k_1\cdots k_n}$ considered
in this section enter into the
construction of these response functions,
and as such this setting provides an intuitive interpretation of them.

\section{The $RA$ basis}
In equilibrium situations one can consider transformations in which
both of the constraints of Eqs.~(\ref{rel1}, \ref{rel2}) translate
into the trivial vanishing of two of the transformed functions
${\widehat D}_{XY}(p)$ of Eq.~(\ref{mat}).
This might be viewed as an attempt to ``economize'' the information
contained in the original propagators $D_{ab}(p)$.
We will consider two classes of such transformations.
We first look at the one that results by demanding
${\widehat D}_{11}(p)$
 vanishes by Eq.~(\ref{rel1}) and
${\widehat D}_{22}(p)$ vanishes by Eq.~(\ref{rel2}). We find
in this case that the transformation automatically leads
to an off-diagonalization in terms of the retarded and advanced
propagators \cite{mich}:
\begin{equation}
{\widehat D}_{\alpha_1\alpha_2}(p)=
\left( \begin{array}{cc}
 {\widehat D}_{RR}(p) & {\widehat D}_{RA}(p)\\
 {\widehat D}_{AR}(p) &  {\widehat D}_{AA}(p)
\end{array}\right)=
\left( \begin{array}{cc}
0 & \Delta_A(p)\\
\Delta_R(p) & 0
\end{array}\right),
\end{equation}
where $\alpha_i= R,\,A$ and,
for the normalization on the right-hand-side to occur,
\begin{equation}
U_{\alpha a}(p)=\frac{1}{a(-p)}\left( \begin{array}{cc}
a(p)a(-p) & -e^{\sigma p_0}a(p)a(-p)   \\
-\eta n(-x) &
 -\eta e^{\sigma p_0}n(x) \end{array}\right),
\label{ur}\end{equation}
with $a(p)$ an arbitrary function.

By Eq.~(\ref{tran}) we then define ``$RA$'' $n$--point functions
${\cal R}_{\alpha_1 \cdots \alpha_n}(p_1,\ldots,p_n)$;
the matrix $V(p)$ of Eq.~(\ref{v}) is,
from Eq.~(\ref{ur}),
\begin{equation}
\left(\begin{array}{cc}
V_{R+}(p) & V_{R-}(p)\\
V_{A+}(p) & V_{A-}(p) \end{array}\right)=
\frac{1}{a(-p)}\left(\begin{array}{cc}
 -\eta n(-x) &
\eta e^{\sigma p_0}n(x) \\
a(p)a(-p) & e^{\sigma p_0}a(p)a(-p)
 \end{array}\right).
\label{ret}\end{equation}
In this notation the non-zero
2--point functions are the self-energies
${\cal R}_{RA}(p_1,p_2)={\cal R}_{AR}^*(p_1,p_2)=\Gamma^{(R)}(p_1;p_2)$.
Using the largest/smallest time equation
relations of Eq.~(\ref{lte}), one finds in general \cite{aur1,mich}
\begin{eqnarray}
& &{\cal R}_{RR\cdots R}(p_1,\ldots,p_n)=0=
{\cal R}_{AA\cdots A}(p_1,\ldots,p_n),\nonumber\\
& &{\cal R}_{RAA\cdots A}(p_1,\ldots,p_n) =
\frac{a(p_2)\cdots a(p_n)}{a(-p_1)}
\Gamma^{(R)}(p_1;p_2,\ldots,p_n),\nonumber\\
& &{\cal R}_{ARR\cdots R}(p_1,\ldots,p_n) = (-1)^n
\frac{a(p_1)n(-x_{p_2})\cdots n(-x_{p_n})}
{n(x_{p_1})a(-p_2)\cdots a(-p_n)}
\Gamma^{(A)}(p_1;p_2,\ldots,p_n),\nonumber\\ & &
\label{genra}\end{eqnarray}
where $\Gamma^{(R)}(p_1;p_2,\ldots,p_n)$
and $\Gamma^{(A)}(p_1;p_2,\ldots,p_n)$
appear in Eq.~(\ref{retadv}).

The $RA$ functions satisfy a complex conjugation relation;
relating $V_{\alpha a}(p)$ to $V_{ {\bar\alpha} {\bar a}}(p)$ and
using Eq.~(\ref{pmcc}), one finds
\begin{eqnarray}
& & {\cal R}^*_{\alpha_1\cdots\alpha_n}(p_1,p_2,\ldots,p_n)= -
 {\cal R}_{\bar{\alpha}_1\cdots\bar{\alpha}_n}(p_1,p_2,\ldots,p_n)
\times\nonumber\\ & &\qquad
\prod_{\alpha_i=R}\frac{n(x_i)}{a(p_i)a(-p_i)}
\prod_{\alpha_i=A}\frac{a(p_i)a(-p_i)}{-n(-x_i)},
\label{racc}\end{eqnarray}
where $\bar\alpha=A,\,R$ for $\alpha=R,\,A$.
This agrees with the results of Refs.~\cite{guer2,mich} for the
choice $a(p)=-\eta n(x)$. We see by Eq.~(\ref{racc}) that it is
not possible to choose
$a(p)$ so as to cancel the factors of the distribution function
$n(x)$ arising in this complex conjugation relation.

Feynman rules for the $RA$ functions consist of internal
retarded or advanced propagators and bare vertices
obtained from Eqs.~(\ref{tran}, \ref{ret}) using the
bare $+/-$ vertices of Eq.~(\ref{vertices}):
\begin{equation}
\frac{ {\widehat g}_{\alpha_1\cdots \alpha_n}(p_1,\ldots,p_n)}
{g(p_1,\ldots,p_n)}=
\prod_{\alpha_i=R} \frac{n(x_i)}{a(-p_i)}
\prod_{\alpha_i=A} a(p_i)
\left[\prod_{\alpha_i=R} e^{x_i}-\prod_{\alpha_i=R}\eta_i\right].
\end{equation}
Thus, as for the full $n$--point functions in Eq.~(\ref{genra}),
bare vertices with all ``$R$'' or all ``$A$'' indices vanish:
${\widehat g}_{RR\cdots R}=0={\widehat g}_{AA\cdots A}$.
The non-vanishing ones satisfy
\begin{equation}
\frac{ {\widehat g}_{\alpha_1\cdots \alpha_n}(-p_1,\ldots,-p_n)}
{{\widehat g}_{\alpha_1\cdots \alpha_n}(p_1,\ldots,p_n)}=
-(-1)^{\#R}\ \frac{ g(-p_1,\ldots,-p_n)}{g(p_1,\ldots,p_n)}
\prod_{\alpha_i=R,A} \frac{a(-p_i)}{a(p_i)},
\label{minusra}\end{equation}
where $\#R$ denotes the number of ``$R$'' indices present.

Two convenient choices of the free parameter $a(p)$ can be
made based on these considerations. One is $a(p)=1$,
which normalizes the bare vertex
${\widehat g}_{RAA\cdots A}$ to the bare $+/\,-$
vertex $g$ of Eq.~(\ref{vertices}). In this case
the factor $\prod a(-p)/a(p)$ in Eq.~(\ref{minusra})
is unity, and the full $n$--point
function ${\cal R}_{RAA\cdots A}$ of Eq.~(\ref{genra}) is normalized
to $\Gamma^{(R)}$. The other choice, as made in Refs.~\cite{aur1,mich},
 is $a(p)=-\eta n(x)$, which normalizes
${\widehat g}_{ARR\cdots R}$ to the bare $+/-$ vertex $g$.
For this choice the factor $\prod a(-p)/a(p)$ in Eq.~(\ref{minusra})
 is $(-1)^n$, and the full $n$--point
function ${\cal R}_{ARR\cdots R}$ of Eq.~(\ref{genra}) is normalized
to $\Gamma^{(A)}$.
\section{The $F\bar F$ basis}
As a second example of using Eqs.~(\ref{rel1}, \ref{rel2})
to eliminate two of the transformed functions
${\widehat D}_{XY}(p)$ in Eq.~(\ref{mat}),
we consider the case where the off-diagonal components
${\widehat D}_{12}(p)$ and ${\widehat D}_{21}(p)$ vanish.
To avoid a singular transformation we find we have to impose,
for example, that ${\widehat D}_{21}(p)$ vanishes by
Eq.~(\ref{rel1}) for $p_0>0$ and by Eq.~(\ref{rel2}) for $p_0<0$,
and {\it vice-versa} for ${\widehat D}_{12}(p)$.
Such a transformation automatically leads to a
diagonalization in terms of the Feynman and anti-Feynman
propagators:
\begin{equation}
{\widehat D}_{f_1 f_2}(p)=
\left( \begin{array}{cc}
 {\widehat D}_{FF}(p) & {\widehat D}_{F{\bar F}}(p)\\
 {\widehat D}_{{\bar F}F}(p) &  {\widehat D}_{{\bar F}{\bar F}}(p)
\end{array}\right)=
\left( \begin{array}{cc}
\Delta_F(p) & 0\\
0 & \Delta_F^*(p)
\end{array}\right),
\end{equation}
where $f_i= F,\,\bar F$ and, for the normalization on the right-hand-side
to occur,
\begin{eqnarray}
& &U_{fa}(p)=\theta(p_0)\left( \begin{array}{cc}
c(p) & -\eta e^{\sigma p_0-x}c(p)   \\
-\eta d(p)& \eta e^{\sigma p_0}d(p) \end{array}\right)
\nonumber\\
& &\qquad\quad+\theta(-p_0)\frac{n(-x)e^{\sigma p_0-x}}
{c(-p)d(-p)}\left( \begin{array}{cc}
e^{-\sigma p_0}d(-p) & - d(-p)   \\
-e^{x-\sigma p_0} c(-p)& \eta c(-p) \end{array}\right),
\label{uf}\end{eqnarray}
where $c(p)$ and $d(p)$ are arbitrary functions.

We next introduce ``$F\bar F$'' $n$--point functions
${\cal F}_{f_1 \cdots f_n}(p_1,\ldots,p_n)$ by Eq.~(\ref{tran});
the matrix $V(p)$ of Eq.~(\ref{v}) is,
from Eq.~(\ref{uf}),
\begin{eqnarray}
& &\left(\begin{array}{cc}
V_{F+}(p) & V_{F-}(p)\\
V_{{\bar F}+}(p) & V_{{\bar F}-}(p) \end{array}\right)=
\theta(p_0)\left( \begin{array}{cc}
c(p) & \eta e^{\sigma p_0-x}c(p)   \\
\eta d(p)& \eta e^{\sigma p_0}d(p) \end{array}\right)
\nonumber\\
& &\quad+\theta(-p_0)\frac{n(-x)e^{\sigma p_0-x}}
{c(-p)d(-p)}\left( \begin{array}{cc}
e^{-\sigma p_0}d(-p) & d(-p)   \\
e^{x-\sigma p_0}c(-p)& \eta c(-p) \end{array}\right).
\label{feyn}\end{eqnarray}
The non-vanishing 2--point functions are then
${\cal F}_{FF}(p_1,p_2)=-{\cal F}_{{\bar F}{\bar F}}^*(p_1,p_2)
=\theta(p_{1_0})\Gamma^{(R)}(p_1;p_2)+
\theta(-p_{1_0})\Gamma^{(A)}(p_1;p_2)$.
It is worth noting that
${\cal F}_{FF}$, not the original amputated 2--point function
$\Gamma_{++}$, determines the pole of the time-ordered
Green function $G_{++}(p)$ of Eq.~(\ref{mat}), and that at
finite temperature these two functions do not
coincide, even at the one--loop level \cite{land,circ}.
A similar lack of equivalence can also be shown for the higher
$n$--point functions ${\cal F}_{FF\cdots F}$ and $\Gamma_{++\cdots+}$.

Feynman rules for the $F\bar F$ functions
consist of the Feynman and anti-Feynman propagators and
bare $F\bar F$ vertices obtained from
Eqs.~(\ref{tran}, \ref{feyn}) using the bare $+/-$
vertices of Eq.~(\ref{vertices}):
\begin{eqnarray}
& & {\widehat g}_{f_1\cdots f_n}(p_1,\ldots,p_n)/g(p_1,\ldots,p_n)=
\nonumber\\
& &\prod_{f_i=F} \left[ \theta(p_{i_0})c(p_i)
+\theta(-p_{i_0})\frac{e^{-x_i}n(-x_i)}{c(-p_i)}\right]
\prod_{f_i={\bar F}} \left[\eta_i \theta(p_{i_0})d(p_i)
+\theta(-p_{i_0})\frac{n(-x_i)}{d(-p_i)} \right]
-\nonumber\\
& &\prod_{f_i=F} \left[ \theta(p_{i_0}) c(p_i)
+\eta_i\theta(-p_{i_0})\frac{n(-x_i)}{c(-p_i)}\right]
\prod_{f_i={\bar F}} \left[ \theta(p_{i_0})e^{x_i}d(p_i)
+\theta(-p_{i_0})\frac{n(-x_i)}{d(-p_i)}\right],
\label{bare}\end{eqnarray}
which satisfy
\begin{eqnarray}
& &\frac{ {\widehat g}_{f_1 \cdots f_n}(-p_1,\ldots,-p_n) }
{{\widehat g}_{\bar{f_1} \cdots \bar{f_n}}(p_1,\ldots,p_n) }=
-\frac{  g(-p_1,\ldots,-p_n) }{g(p_1,\ldots,p_n)}\times
\nonumber\\
& &\quad \prod_{f_i=F,{\bar F}}
\left[\theta(p_{i_0}) \frac{n(x_i)}{c(p_i)d(p_i)}
 +\theta(-p_{i_0}) \frac{c(-p_i)d(-p_i)}{n(-x_i)} \right].
\label{minus}\end{eqnarray}

Two considerations illustrate the relative effects
of different choices of the arbitrary coefficients $c(p)$ and $d(p)$
in this basis.
One is the normalization of the
$F\bar F$ $n$--point functions, for which we have, for example,
\begin{eqnarray}
& & \theta(k_0)\theta(-l_0)\cdots\theta(-r_0)
{\cal F}_{FF\cdots F}(k,l,\ldots,r)=\nonumber\\
& & =\theta(k_0)\theta(-l_0)\cdots\theta(-r_0)
\left[\frac{c(k)n(-x_l)\cdots n(-x_r)}
{n(x_k)c(-l)\cdots c(-r)}\right]
\Gamma^{(R)}(k;l,\ldots,r), \nonumber\\
& & \theta(k_0)\theta(-l_0)\cdots\theta(-q_0)\theta(r_0)
{\cal F}_{FF\cdots F{\bar F}}(k,l,\ldots,q,r)=\nonumber\\ & &=
\theta(k_0)\theta(-l_0)\cdots\theta(-q_0)\theta(r_0)
\left[\frac{c(k)n(-x_l)\cdots n(-x_q) d(r)}
{n(x_k)c(-l)\cdots c(-q)\eta_r e^{-x_r}}\right]
\Gamma^{(R)}(k;l,\ldots,q,r),\nonumber\\
& & \theta(-k_0)\theta(l_0)\cdots\theta(r_0)
{\cal F}_{{\bar F}{\bar F}\cdots {\bar F}}(k,l,\ldots,r)=\nonumber\\
& & =-\theta(-k_0)\theta(l_0)\cdots\theta(r_0)
\left[\frac{d(l)\cdots d(r)}{d(-k)}\right]
\Gamma^{(R)}(k;l,\ldots,r),\nonumber\\
& & \theta(-k_0)\theta(l_0)\cdots\theta(r_0)
{\cal F}_{FF\cdots F}(k,l,\ldots,r)=\nonumber\\
& & =\theta(-k_0)\theta(l_0)\cdots\theta(r_0)
\left[\frac{c(l)\cdots c(r)}{c(-k)}\right]
\Gamma^{(A)}(k;l,\ldots,r), \nonumber\\
& & \theta(-k_0)\theta(l_0)\cdots\theta(q_0)\theta(-r_0)
{\cal F}_{FF\cdots F{\bar F}}(k,l,\ldots,q,r)=\nonumber\\ & &=
\theta(-k_0)\theta(l_0)\cdots\theta(q_0)\theta(-r_0)
\left[\frac{ c(l)\cdots c(q)n(-x_r)}
{c(-k)d(-r)}\right]
\Gamma^{(A)}(k;l,\ldots,q,r),\nonumber\\
& & \theta(k_0)\theta(-l_0)\cdots\theta(-r_0)
{\cal F}_{{\bar F}{\bar F}\cdots {\bar F}}(k,l,\ldots,r)=\nonumber\\
& & =-\theta(k_0)\theta(-l_0)\cdots\theta(-r_0)
\left[\frac{d(k)n(-x_l)\cdots n(-x_r)}
{n(x_k)d(-l)\cdots d(-r)}\right]
\Gamma^{(A)}(k;l,\ldots,r),
\label{norm}\end{eqnarray}
where $\Gamma^{(R)}(k;l,\ldots,r)$ and $\Gamma^{(A)}(k;l,\ldots,r)$
are defined in Eq.~(\ref{retadv}).
The other consideration is complex conjugation:
relating $V_{f a}$ to $V_{ {\bar f}{\bar a} }$ from Eq.~(\ref{feyn})
and using Eq.~(\ref{pmcc}) one can show
\begin{eqnarray}
& &{\cal F}_{f_1\cdots f_n}^*(p_1,\ldots,p_n)=
-{\cal F}_{{\bar f_1}\cdots {\bar f_n}}(p_1,\ldots,p_n)\times
\nonumber\\
& & \prod_{f_i=F}\left[
\theta(p_{i_0})\frac{c(p_i)}{d(p_i)}+
\theta(-p_{i_0})\frac{d(-p_i)}{c(-p_i)} \right]
 \prod_{f_i={\bar F}}\left[
\theta(p_{i_0})\frac{d(p_i)}{c(p_i)}+
\theta(-p_{i_0})\frac{c(-p_i)}{d(-p_i)} \right]
\eta_i e^{x_i},\nonumber\\ & &
\label{comp}\end{eqnarray}
where ${\bar f}={\bar F},\,F$ for $f=F,\,{\bar F}$.

We consider now in the preceding context two choices of the
parameters $c(p)$ and $d(p)$. For the first one, we begin by
noting that
the ``symmetrical'' or ``unitary'' formulation of thermo field
dynamics is recovered by the choices
\begin{eqnarray}
& &\sigma=\beta/2, \nonumber\\
& &c(p)=e^{x/2}\sqrt{n(x)},\nonumber\\
& &d(p)=\eta e^{-\beta\mu/2}\sqrt{n(x)},
\label{tfd}\end{eqnarray}
in the transformation matrix
of Eq.~(\ref{feyn}) \cite{land,um,henn}. Of course,
potential problems exist for massive bosons in this when
taking the square root of the distribution function.
However, the $\theta$ functions of energy which accompany
$c(p)$ and $d(p)$ in general guarantee that this square root
remains real except in the case of finite
chemical potential with
$0<(p_0/\mu)<1$, where a separate analysis in this and the following
is required \cite{land,aur1}.
Guided by the values
of Eq.~(\ref{tfd}) we then consider, for arbitrary
$\sigma$, the choices
\begin{eqnarray}
& &c(p)=e^{ x/2}\sqrt{n(x)},\nonumber\\
& &d(p)=\eta \sqrt{n(x)}.
\end{eqnarray}
In this case the bare vertices of Eq.~(\ref{bare}) are
\begin{equation}
\frac{ {\widehat g}_{f_1\cdots f_n}(p_1,\ldots,p_n)}{g(p_1,\ldots,p_n)}=
\prod_{f_i=F} C[p_i]
\prod_{f_i={\bar F}}\eta_i S[p_i]-
\prod_{f_i=F} S[p_i] \prod_{f_i={\bar F}} C[p_i],
\end{equation}
which satisfy
\begin{equation}
\frac{  {\widehat g}_{f_1 \cdots f_n}(-p_1,\ldots,-p_n) }
{{\widehat g}_{\bar{f_1} \cdots \bar{f_n}}(p_1,\ldots,p_n) }=
-\frac{  g(-p_1,\ldots,-p_n) }{g(p_1,\ldots,p_n)},
\end{equation}
and where
\begin{eqnarray}
& & C[p]=\theta(p_0)e^{x/2}\sqrt{n(x)} +
\theta(-p_0)e^{-x/2}\sqrt{n(-x)},\nonumber\\
& & S[p]=\theta(p_0)\sqrt{n(x)} +
\eta \theta(-p_0)\sqrt{n(-x)},
\end{eqnarray}
which obey $C^2[p]-\eta S^2[p]=1$.
This choice of parameters leads to the following prefactors
in the normalization relations of Eq.~(\ref{norm}):
\begin{eqnarray}
& &\frac{c(k)n(-x_l)\cdots n(-x_r)}
{n(x_k)c(-l)\cdots c(-r)}=
\sqrt{\frac{n(-x_l)\cdots n(-x_r)}{n(x_k)}}, \nonumber\\
& & \frac{c(k)n(-x_l)\cdots n(-x_q) d(r)}
{n(x_k)c(-l)\cdots c(-q)\eta_r e^{-x_r}}=
e^{x_r/2}
\sqrt{\frac{n(-x_l)\cdots n(-x_q)n(x_r)}{n(x_k)}},\nonumber\\
& &\frac{d(l)\cdots d(r)}{d(-k)}=
\sqrt{\frac{n(x_l)\cdots n(x_r)}{n(-x_k)}},\nonumber\\
& & \frac{c(l)\cdots c(r)}{c(-k)}=
\sqrt{\frac{n(x_l)\cdots n(x_r)}{n(-x_k)}}, \nonumber\\
& &\frac{ c(l)\cdots c(q)n(-x_r)}
{c(-k)d(-r)}=
\eta_r e^{- x_r/2}
\sqrt{\frac{n(x_l)\cdots n(x_q)n(-x_r)}{n(-x_k)}},\nonumber\\
& &\frac{d(k)n(-x_l)\cdots n(-x_r)}
{n(x_k)d(-l)\cdots d(-r)}=
\sqrt{\frac{n(-x_l)\cdots n(-x_r)}{n(x_k)}},
\end{eqnarray}
while the complex conjugation relation of Eq.~(\ref{comp}) is
\begin{equation}
{\cal F}_{f_1\cdots f_n}^*(k,\ldots,r)=
-{\cal F}_{{\bar f_1}\cdots {\bar f_n}}(k,\ldots,r)
 \prod_{f_i={\bar F}}
\eta_i .
\end{equation}

As a second choice of the parameters $c(p)$ and $d(p)$, we consider
\begin{eqnarray}
& & c(p)=e^x n(x),\nonumber\\
& & d(p)=\eta.
\end{eqnarray}
This choice leads to bare vertices of Eq.~(\ref{bare}) of the form
\begin{eqnarray}
& & \frac{{\widehat g}_{f_1\cdots f_n}(p_1,\ldots,p_n)}{g(p_1,\ldots,p_n)}
=\prod_{f_i=F} \left[1+\eta_i \theta(p_{i_0})n(x_i)\right]
\prod_{f_i={\bar F}} \left[\theta(p_{i_0})
+ \eta_i \theta(-p_{i_0}) n(-x_i) \right]
-\nonumber\\
& &\quad\prod_{f_i=F} \left[
\eta_i\theta(p_{i_0})n(x_i)+\theta(-p_{i_0}) \right]
\prod_{f_i={\bar F}} \left[
1+\eta_i\theta(-p_{i_0})n(-x_i)\right],
\end{eqnarray}
which satisfy
\begin{equation}
\frac{  {\widehat g}_{f_1 \cdots f_n}(-p_1,\ldots,-p_n) }
{{\widehat g}_{\bar{f_1} \cdots \bar{f_n}}(p_1,\ldots,p_n) }=
-\frac{  g(-p_1,\ldots,-p_n) }{g(p_1,\ldots,p_n)}.
\end{equation}
In this case the prefactors in the
normalization conditions of Eq.~(\ref{norm})
become
\begin{eqnarray}
& &\frac{c(k)n(-x_l)\cdots n(-x_r)}
{n(x_k)c(-l)\cdots c(-r)}=1, \nonumber\\
& & \frac{c(k)n(-x_l)\cdots n(-x_q) d(r)}
{n(x_k)c(-l)\cdots c(-q)\eta_r e^{-x_r}}=1,\nonumber\\
& &\frac{d(l)\cdots d(r)}{d(-k)}=1,\nonumber\\
& &\frac{c(l)\cdots c(r)}{c(-k)}=
\frac{n(x_l)\cdots n(x_r)}{n(-x_k)}, \nonumber\\
& &\frac{ c(l)\cdots c(q)n(-x_r)}
{c(-k)d(-r)}=
\eta_r e^{-x_r}\frac{n(x_l)\cdots n(x_q)n(-x_r)}{n(-x_k)},\nonumber\\
& &\frac{d(k)n(-x_l)\cdots n(-x_r)}
{n(x_k)d(-l)\cdots d(-r)}=
\frac{n(-x_l)\cdots n(-x_r)}{n(x_k)},
\end{eqnarray}
while the complex conjugation relation of Eq.~(\ref{comp}) is
\begin{eqnarray}
& &{\cal F}_{f_1\cdots f_n}^*(p_1,\ldots,p_n)=
-{\cal F}_{{\bar f_1}\cdots {\bar f_n}}(p_1,\ldots,p_n)\times
\nonumber\\
& & \prod_{f_i=F}\left[ \theta(p_{i_0}) n(x_i)
+\frac{\theta(-p_{i_0})}{n(-x_i)} \right]
\prod_{f_i={\bar F}}\left[ \frac{\theta(p_{i_0})}{n(x_i)}+
\theta(-p_{i_0}) n(-x_i) \right]\eta_i e^{-x_i}.
\end{eqnarray}

One can also consider parameters
$c(p)$ and $d(p)$ that represent choices that interpolate
between the two competing features of normalization
and complex conjugation. For example, the choice
$c(p)=e^{x/2}$ and $d(p)=1$
leads to a simple complex conjugation relation of Eq.~(\ref{comp})
but also to a somewhat
more involved set of normalization relations of Eq.~(\ref{norm});
this has the advantage that no square roots of distribution
functions need be taken, but at the expense of having a
more complicated form of Eq.~(\ref{minus}) relating the
bare vertex ${\widehat g}_{f_1 \cdots f_n}(-p_1,\ldots,-p_n)$
to ${\widehat g}_{{\bar f_1} \cdots {\bar f_n}}(p_1,\ldots,p_n)$.

\section{Comparisons of the bases}

In this section we look at the transformations between the
three bases considered here. As well as comparing their
respective structures, this will allow us to see how relations
in one basis translate into relations in another basis.
We first consider the transformation between the
$RA$ functions and the Keldysh-like 1/2 functions.
Writing this relation as
\begin{equation}
{\cal R}_{\alpha_1 \cdots \alpha_n}(p_1,\ldots,p_n) =
T_{\alpha_1 k_1}(p_1)\cdots T_{\alpha_n k_n}(p_n)
{\cal K}_{k_1\cdots k_n}(p_1,\ldots,p_n)
\end{equation}
we find, for the choice $b(p)=1/\sqrt{2}$ and $a(p)=1$ in
Eqs.~(\ref{kelv}, \ref{ret}), the matrix
\begin{equation}
\left(\begin{array}{cc}
T_{R1}(p) & T_{R2}(p)\\
T_{A1}(p) & T_{A2}(p) \end{array}\right)
=\left(\begin{array}{cc}
\frac{1}{2}\coth^\eta (x/2) & 1\\
1 & 0 \end{array}\right).
\label{trankr}\end{equation}
With this transformation, it is seen that the identity
${\cal R}_{AA\cdots A}=0$ in the $RA$ basis
translates in the Keldysh basis to
${\cal K}_{11\cdots 1}=0$. The other identity in the $RA$ basis,
${\cal R}_{RR\cdots R}=0$, gives a relation amongst various
${\cal K}_{k_1\cdots k_n}$ functions. For the
2--point function this relation is the fluctuation-dissipation
theorem of the last relation of Eq.~(\ref{con}), while for
higher $n$--point functions such a relation can be used to
construct what have been advocated as generalizations of the
fluctuation-dissipation theorem \cite{chou}.
One can also see using Eq.~(\ref{trankr}) that the $n$--point
retarded functions in the two bases are related as
${\cal R}_{RAA\cdots A}={\cal K}_{211\cdots 1}$.

It is also of interest to examine other ``mixed'' functions, as
for example
\begin{eqnarray}
& &{\cal R}_{RRAA\cdots A}(p_1,p_2,\ldots,p_n)=
{\cal K}_{2211\cdots 1}(p_1,p_2,\ldots,p_n)+\nonumber\\
& &\qquad
+\frac{1}{2}\coth^\eta(x_{p_1}/2){\cal K}_{1211\cdots 1}(p_1,p_2,\ldots,p_n)
+\nonumber\\ & &\qquad
+\frac{1}{2}\coth^\eta(x_{p_2}/2){\cal K}_{2111\cdots 1}(p_1,p_2,\ldots,p_n).
\label{mix}\end{eqnarray}
Consider first the 3--point functions. Recent studies
have examined the relation between $n$--point functions of
real-time formalisms and the analytically continued
functions $\Gamma^{(n)}(z_1,\ldots,z_n)$ of the imaginary-time formalism,
where initially all energy components of $z_i$ are
in the discrete set of Matsubara frequencies on the imaginary axis.
For the retarded products the relation
is \cite{evans1,evans2,nieg,guer2,aur1}
\begin{eqnarray}
& & {\cal R}_{RAA\cdots A}(p_1,p_2,\ldots,p_n)=
{\cal K}_{211\cdots 1}(p_1,p_2,\ldots,p_n)=\nonumber\\ & &\quad
=\Gamma^{(n)}(z_1\to p_{1_0}+i(n-1)\varepsilon, z_2\to p_{2_0}-i\varepsilon,
\ldots, z_n\to p_{n_0}-i\varepsilon).
\label{raim}\end{eqnarray}
Together with the complex conjugation relation of Eq.~(\ref{racc}),
this allows us to infer from Eq.~(\ref{mix}) the relation between
the 3--point response function ${\cal K}_{221}$ and various analytic
continuations of the imaginary-time functions.
Such a relation to the imaginary-time functions can also be
worked out using Eq.~(\ref{mix})
for the 4--point response function ${\cal K}_{2211}$. For
this we need, as well as Eq.~(\ref{raim}), the relation
between, for example,
the mixed $RA$ function ${\cal R}_{AARR}$ and various
analytic continuations of the 4--point imaginary-time
function \cite{guer2,tay}.

We also note that
Eq.~(\ref{mix}) can provide an interpretation of the ``mixed''
$RA$ functions in terms of response functions in the
Keldysh basis. As a result, certain
causal properties (in time)
can be inferred. For example, since the response
relations of Eq.~(\ref{resp}) are causal in nature, we conclude that
each of the functions $G_{1222\cdots 2}(t_1,t_2,\ldots)$,
$G_{2122\cdots 2}(t_1,t_2,\ldots)$, and
$G_{1122\cdots 2}(t_1,t_2,\ldots)$ should only have
$t_1$ or $t_2$ as the largest time \cite{chou,resp}.
By a similar analysis for the non-amputated
mixed functions $G_{RRAA\cdots A}$ as that leading
to Eq.~(\ref{mix}), one could then deduce the corresponding
causal properties for these mixed $RA$ Green functions.

We next consider the transformation between the $F\bar F$ functions
and those of the $RA$ basis. Writing this as
\begin{equation}
{\cal R}_{\alpha_1 \cdots \alpha_n}(p_1,\ldots,p_n) =
T_{\alpha_1 f_1}(p_1)\cdots T_{\alpha_n f_n}(p_n)
{\cal F}_{f_1\cdots f_n}(p_1,\ldots,p_n)
\end{equation}
we find, for the choices $a(p)=1$ and $c(p)=e^x n(x),\,
d(p)=\eta$ in Eqs.~(\ref{ret}, \ref{feyn}), the matrix
\begin{equation}
\left(\begin{array}{cc}
T_{RF}(p) & T_{R\bar F}(p)\\
T_{AF}(p) & T_{A{\bar F}}(p) \end{array}\right)
=\theta(p_0) \left( \begin{array}{cc}
1 & 0  \\
0 & 1 \end{array}\right)
+\theta(-p_0)
\left( \begin{array}{cc}
0& -1  \\
1 & 0 \end{array}\right).
\label{trans}\end{equation}
 From this one sees that an $F$ index corresponds to an $R$ index
for positive energies and to an $A$ index for negative energies,
and {\it vice-versa} for an $\bar F$ index.
As well, with this relation one can translate the identities
${\cal R}_{RR\cdots R}=0={\cal R}_{AA\cdots A}$
of the $RA$ basis into the following set of relations in the
$F\bar F$ basis:
\begin{eqnarray}
& &\theta(k_0)\cdots\theta(m_0)\theta(-n_0)\cdots\theta(-r_0)
{\cal F}_{F\cdots F\,{\bar F}\cdots{\bar F}}(k,\ldots,m,n,\ldots,r)=0,
\nonumber\\
& &\theta(-k_0)\cdots\theta(-m_0)\theta(n_0)\cdots\theta(r_0)
{\cal F}_{F\cdots F\,{\bar F}\cdots{\bar F}}(k,\ldots,m,n,\ldots,r)=0.
\label{causal}\end{eqnarray}
These relations have
a direct causality interpretion: if an
$F$ index represents a particle ($p_0>0$)
travelling forward in time and an antiparticle ($p_0<0$) travelling
backwards in time, and an $\bar F$ index represents the
time-reversed situation,
then Eq.~(\ref{causal}), together with the complex conjugation
relation of Eq.~(\ref{comp}), states that one cannot have particles
and antiparticles either all disappearing into the vacuum nor all
appearing out of the vacuum. Such an interpretation also holds for
arbitrary choices of the parameters $a(p)$, $c(p)$, and $d(p)$.

With the inverse of the transformation of Eq.~(\ref{trans})
we can, given relations such as Eq.~(\ref{raim})
relating the $RA$ functions to the
analytically continued imaginary-time functions
$\Gamma^{(n)}$, examine how
the $F\bar F$ functions are related to the imaginary-time
functions.
To simplify the notation, from now on we omit from
 $\Gamma^{(n)}(z_1,\ldots,z_n)$ the last complex argument $z_n$,
 which by energy-momentum conservation is minus the sum of the others.
We first consider, with $p+q=0$, the 2--point function
${\cal F}_{FF}(p,q)=-{\cal F}_{{\bar F}{\bar F}}^*(p,q)$,
for which we find
\begin{eqnarray}
& &{\cal F}_{FF}(p,q)
=\theta(p_0)\Gamma^{(R)}(p;q)+\theta(-p_0)\Gamma^{(A)}(p;q)
=\nonumber\\
& &\ =\theta(p_0)\Gamma^{(2)}(z\to p_0+i\varepsilon)+
\theta(-p_0)\Gamma^{(2)}(z\to p_0-i\varepsilon)=
\Gamma^{(2)}(z\to p_0+i\varepsilon p_0),
\nonumber\\ & &\label{twopoint}\end{eqnarray}
as expected \cite{land}. The other two 2--point functions
${\cal F}_{F\bar F}(p,q)\sim {\cal F}^*_{ {\bar F} F}(p,q)$
vanish identically, as can be shown
by multiplying ${\cal F}_{F\bar F}(p,q)$ by $1=\theta(p_0)+\theta(-p_0)$
and then using the causality relations of Eq.~(\ref{causal}).

We next consider the 3--point function. With
$p+q+r=0$, we first examine
${\cal F}_{FFF}(p,q,r)$, which using
Eqs.~(\ref{racc}, \ref{raim}, \ref{trans})
can be written as
\begin{eqnarray}
& &{\cal F}_{FFF}(p,q,r)=
\theta(-p_0)\theta(-q_0)
\Gamma^{(3)}(z_1\to p_0-i\varepsilon, z_2\to q_0-i\varepsilon)\nonumber\\
& & \quad+\theta(-p_0)\theta(-r_0)
\Gamma^{(3)}(z_1\to p_0-i\varepsilon, z_2\to q_0+2i\varepsilon)\nonumber\\
& & \quad+\theta(-q_0)\theta(-r_0)
\Gamma^{(3)}(z_1\to p_0+2i\varepsilon, z_2\to q_0-i\varepsilon)\nonumber\\
& & \quad+\theta(p_0)\theta(q_0)
\frac{n(x_p)n(x_q)}{n(-x_r)}
\Gamma^{(3)}(z_1\to p_0+i\varepsilon, z_2\to q_0+i\varepsilon)\nonumber\\
& & \quad+\theta(p_0)\theta(r_0)
\frac{n(x_p)n(x_r)}{ n(-x_q)}
\Gamma^{(3)}(z_1\to p_0+i\varepsilon, z_2\to q_0-2i\varepsilon)\nonumber\\
& & \quad+\theta(q_0)\theta(r_0)
\frac{ n(x_q)n(x_r)}{n(-x_p)}
\Gamma^{(3)}(z_1\to p_0-2i\varepsilon, z_2\to q_0+i\varepsilon).
\label{FFF}\end{eqnarray}
In this form ${\cal F}_{FFF}$ appears to be a linear combination
of the three possible analytic continuations of the 3--point
imaginary-time function and their complex conjugates,
but the presence of the $\theta$ functions
allows the replacement in Eq.~(\ref{FFF}) of all the
$\Gamma^{(3)}(z_1,z_2)$ functions by a single
analytically continued function,
$\Gamma^{(3)}(z_1\to p_0+i\varepsilon p_0, z_2\to q_0+i\varepsilon q_0)$
\cite{rolf,three}.
Up to a normalization factor, then, this result
could have been anticipated from the 2--point function
${\cal F}_{FF}(p,q)$ of Eq.~(\ref{twopoint}).
In fact, a similar analysis for the $n$--point
function ${\cal F}_{F\cdots F}(p_1,\ldots,p_n)$ indicates
that this ``Feynman'' function in general can be written in
terms of a single analytically continued imaginary-time function
with $z_i\to p_{i_0}+i\varepsilon p_{i_0}$.

One can also analyze the ``mixed'' $F\bar F$
functions in a similar way.
For example, for the 3--point function
${\cal F}_{FF\bar F}(p,q,r)$ we find
\begin{eqnarray}
& &{\cal F}_{FF\bar F}(p,q,r)=
\theta(q_0)\theta(r_0)
\Gamma^{(3)}(z_1\to p_0-i\varepsilon, z_2\to q_0+2i\varepsilon)\nonumber\\
& & \quad+\theta(p_0)\theta(r_0)
\Gamma^{(3)}(z_1\to p_0+2i\varepsilon, z_2\to q_0-i\varepsilon)\nonumber\\
& & \quad-\theta(-q_0)\theta(-r_0)
\frac{n(x_p)n(x_r)}{ n(-x_q)}
\Gamma^{(3)}(z_1\to p_0+i\varepsilon, z_2\to q_0-2i\varepsilon)\nonumber\\
& & \quad-\theta(-p_0)\theta(-r_0)
\frac{ n(x_q)n(x_r)}{n(-x_p)}
\Gamma^{(3)}(z_1\to p_0-2i\varepsilon, z_2\to q_0+i\varepsilon).
\label{FFbF}\end{eqnarray}
However, unlike the ``pure'' Feynman $n$--point function ${\cal F}_{FF\cdots
F}$,
there does not appear to be any simple interpretation of these
mixed $F\bar F$ functions in terms of single analytic continuations
of the imaginary-time functions.

Let us compare the number of independent functions present
in all three bases. Taking into account the complex conjugation
relation of Eq.~(\ref{racc}) in the $RA$ basis and Eq.~(\ref{comp})
in the $F\bar F$ basis, as well as
Eq.~(\ref{con}) for the Keldysh basis,
all three bases are seen to contain one independent
 2--point function, which
can be taken as the retarded function ${\cal R}_{RA}={\cal K}_{21}$
or the Feynman function ${\cal F}_{FF}$.
For the 3--point function the
$RA$ basis has three independent functions, which could be taken as the
three retarded functions ${\cal R}_{RAA}$, ${\cal R}_{ARA}$, and
${\cal R}_{AAR}$. The Keldysh basis contains these
three retarded functions --
${\cal K}_{211}$, ${\cal K}_{121}$, and ${\cal K}_{112}$ -- but also four
additional functions: ${\cal K}_{222}$, ${\cal K}_{122}$,
${\cal K}_{212}$, and ${\cal K}_{221}$. However,
relations such as Eq.~(\ref{mix}) and that following from ${\cal R}_{RRR}=0$
in Eq.~(\ref{trankr}) shows that these latter four can be
written as linear combinations of the retarded functions and their
complex conjugates. Similarly, taking into account
complex conjugation, the $F\bar F$ basis appears to
have four 3--point functions, which could be taken as ${\cal F}_{FFF}$,
${\cal F}_{{\bar F}FF}$, ${\cal F}_{F{\bar F}F}$, and ${\cal F}_{FF{\bar F}}$.
However, relations such as Eqs.~(\ref{FFF}, \ref{FFbF}) show that
all four can be written in terms of the three retarded products and
their complex conjugates. Alternatively, recalling that
these ``mixed'' $F\bar F$ functions are
constrained by the relations of Eq.~(\ref{causal}) forcing
energy to flow in only certain directions, one could in this
context count as an independent function each
distinct orientation of the flow of external energy. In this
way the $RA$ basis has $3\times 6=18$ such independent oriented functions,
while the $F\bar F$ basis also has $1\times 6 + 3\times 4=18$
independent oriented functions. One thus concludes that all three bases
contain the same number of independent 3--point functions.

Various examples could serve to compare the convenience of these
bases in practical calculations.
A simple and illustrative such example is the
one-loop self-energy diagram of Eq.~(\ref{selfenergy})
with a bare $+/-$ coupling constant $g$ of Eq.~(\ref{vertices})
independent of momentum. The corresponding example in the
imaginary-time formalism using the retarded
analytic continuation $z\to p_0+i\varepsilon$ has been
 explored by Weldon \cite{wel}.
Let us then find from Eq.~(\ref{selfenergy}) the
retarded self-energy. We first consider the Keldysh and $RA$ bases,
for which we calculate ${\cal K}_{21}(p,-p)={\cal R}_{RA}(p,-p)$.
In the Keldysh basis the non-vanishing bare vertices
needed are ${\widehat g}_{211}$ and
permutations of its indices, since ${\widehat D}_{22}(p)=0$, and
so of all the possible terms in Eq.~(\ref{selfenergy}) only
two survive. In the $RA$ basis the relevant non-vanishing
bare vertices are ${\widehat g}_{RAA}$
and ${\widehat g}_{ARR}$ and permutations of their indices, since
${\widehat D}_{RR}(p)=0={\widehat D}_{AA}(p)$.
Three possible contributions then remain
in Eq.~(\ref{selfenergy}), but two of these
disappear upon integration over $k_0$ because
of the presence of terms like $\Delta_R(k)\Delta_R(q)$ or
$\Delta_A(k)\Delta_A(q)$ with poles in the same half-plane \cite{aur1}.
In both bases one thus obtains, with $q=k-p$ and neglecting any
$\gamma$-matrix structure,
\begin{eqnarray}
& &{\cal K}_{21}(p,-p)={\cal R}_{RA}(p,-p)=\nonumber\\
& &\qquad=
-ig^2\int {d^4k\over (2\pi)^4}\left\{
\Delta_R(k)\left[ \frac{1}{2}+\eta_q n(x_q)\right]
\left[ \Delta_R(q)-\Delta_A(q)\right]+ \right. \nonumber\\
& &\qquad\qquad+ \left.
\Delta_A(q)\left[ \frac{1}{2}+\eta_k n(x_k)\right]
\left[ \Delta_R(k)-\Delta_A(k)\right] \right\},
\label{result}\end{eqnarray}
which agrees with previous results \cite{aur1,wel,guer1}.

In this calculation the choice of free parameters
$b(p)=1$ or $1/\sqrt{2}$ in the Keldysh basis and
$a(p)=1$ or $-\eta n(x)$ in the $RA$ basis are
easiest to use. The form of the result of Eq.~(\ref{result})
is convenient for doing the $k_0$ integration,
since for free propagators $\Delta_R(k)-\Delta_A(k)=
2\pi\varepsilon(k_0)\delta(k^2-m^2)$. Such a form
also arises for higher $n$--point functions at the one-loop
level and, with suitable modifications, also at higher
loop orders \cite{aur1}. One might have anticipated that
the $RA$ basis would be significantly easier to use
compared to the Keldysh basis: there
are only two $RA$ propagators $\Delta_R$ and $\Delta_A$, whereas
the Keldysh basis also contains the dependent
combination $\Delta_S\sim \coth^\eta(x/2) [\Delta_R-\Delta_A]$,
and also two $RA$ $n$--point functions
${\cal R}_{RR\cdots R}$ and ${\cal R}_{AA\cdots A}$
vanish identically in general, while
in the Keldysh basis only ${\cal K}_{11\cdots 1}$ does so. These
features of the $RA$ basis are certainly an advantage at
higher loop orders. However,
the vanishing of the bare Keldysh vertices with an even number
of ``2'' indices, together with the particular form of $\Delta_S$
which arises directly in the result of Eq.~(\ref{result}),
simplifies calculations in the Keldysh basis considerably.
At least to lower orders, then, the use of
the Keldysh basis and the $RA$ basis as outlined here
are comparable in difficulty.

In the $F\bar F$ basis the
retarded self-energy could be found as $\theta(p_0){\cal F}_{FF}(p,-p)$.
To show the final result of this calculation for Eq.~(\ref{selfenergy})
has the form as Eq.~(\ref{result}) is somewhat tedious.
An easier problem is to compute
the imaginary part of this self-energy, which is
discussed in detail by Weldon in the imaginary-time formalism \cite{wel}.
When this is found using
Eq.~(\ref{selfenergy}) in the $F\bar F$ basis, terms such as
\begin{eqnarray}
& & \int dk_0\ f(k_0,q_0)\left[ \Delta_F(k)\Delta_F(q)
+ \Delta_F^*(k)\Delta_F^*(q)\right]=\nonumber\\
& &\quad= \int dk_0\ f(q_0,r_0)\left[ \Delta^+(k)\Delta^-(q)
+ \Delta^-(k)\Delta^+(q)\right],\nonumber\\
& & \int dk_0\ g(k_0,q_0)\left[ \Delta_F(k)\Delta_F^*(q)
+ \Delta_F^*(k)\Delta_F(q)\right]=\nonumber\\
& &\quad= \int dk_0\ g(k_0,q_0)\left[ \Delta^+(k)\Delta^+(q)
+ \Delta^-(k)\Delta^-(q)\right],
\label{pm}\end{eqnarray}
arise \cite{vel}, where, for example,
 $\Delta^\pm(p)=2\pi\theta(\pm p_0)\delta(p^2-m^2)$ for bosons.
Thus, $\theta$ functions of energy automatically appear which,
when applied to the vertices
${\widehat g}_{fgh}(p,q,r)$, lead to simplifications due to the normalization
and complex conjugation relations described previously
for the $F\bar F$ functions.
The result in the end, of course,
agrees with the corresponding one of Weldon and others
\cite{aur1,circ,wel,guer1}.

For this particular calculation of the imaginary part
the choice of parameters $c(p)=e^x n(x)$ and
$d(p)=\eta$ in the $F\bar F$ basis
is somewhat easier to use. However, in this example
other methods such as the imaginary-time formalism or the
$RA$ or Keldysh bases are simpler,
although this may not be so in more
complex situations. In this regard, we remark that the steps used in this
calculation are suggestive of the Cutkosky or cutting rules
for the imaginary parts of graphs
as derived in the approach of 't Hooft and
Veltman \cite{vel}. These rules at finite temperature
using the $+/-$ $D_{ab}$ propagators
 have been discussed in Refs.~\cite{circ,corr,nak,jeon},
and it is straightforward to derive the corresponding rules for the
$F\bar F$ basis. A feature worth noting about such
a derivation is that, because the zero temperature
Feynman propagator and its complex conjugate are involved,
the ``non-cuttable'' graphs that arise beyond the one-loop
level at finite temperature when the $D_{ab}$ propagators are used will
not be present in the $F\bar F$ rules;
for this the ``causality'' relations of Eq.~(\ref{causal})
play a crucial role.

\section{Discussion}
In many instances physical quantities are more naturally
expressed in terms of Green functions other than
the usual time-ordered products. At finite temperature it is
important to be aware of which function is needed, as
differences between the various types of products is more
pronounced than at zero temperature \cite{evans1,evans2}.
Although in principle it is possible to extract all such information
from the original $D_{ab}$ basis of the real-time formalism,
in practice this may not be the most convenient method, especially in
light of the constraints of Eqs.~(\ref{rel1}, \ref{rel2}) among
the components of $D_{ab}$. For this reason formalisms such
as the Keldysh and Retarded/Advanced bases have been constructed which
involve explicitly, in particular, the retarded propagator.
The transformation of Eq.~(\ref{mat}) from the $D_{ab}$ basis to a
generic transformed ${\widehat D}_{XY}$ basis provides a unifying framework
for these various real-time formalisms, in that the difference between
them can be attributed to the different ways the constraints of
Eqs.~(\ref{rel1}, \ref{rel2}) are implemented.

We have studied here three classes of such transformations -- the Keldysh
basis, the retarded/advanced $RA$ basis, and a Feynman-like $F\bar F$
basis -- which in a sense are attempts to implement the
constraints of Eqs.~(\ref{rel1}, \ref{rel2}) ``economically''.
By means of the transformation between the bases we have also seen
how information in one basis can be translated into the equivalent
information in another basis; this enabled us, for example, to examine
the relation between various real-time functions and certain analytically
continued imaginary-time functions. Finally, we compared some
aspects of these bases as they arise in practice.
Although some of the relations in the $F\bar F$
basis such as complex
conjugation can be chosen to be simpler than the corresponding
ones in the $RA$ basis, it appears in general that the conditions
of Eqs.~(\ref{rel1}, \ref{rel2})
are implemented more efficiently in actual
calculations in the $RA$ basis  \cite{aur1,aur2,aur3,mich2}, or even,
comparable at least to
lower orders, in the Keldysh basis. Thus, for practical
purposes, it appears that of these three bases
the $RA$ or perhaps the Keldysh basis is easier to use
in general, although this could depend on the particular application
and may also be a matter of taste.
\vspace*{20pt}\par\noindent
{\bf Acknowledgments}\vspace*{8pt}\\
We thank P.~Aurenche, T.~Evans, and F.~Guerin for
fruitful discussions. One of us (RK) also thanks the hospitality of
the Institute for Theoretical Physics at the University of California
at Santa Barbara, where part of this work was done.
This research was supported in part by
the Natural Sciences and Engineering Research Council of Canada,
the National Science Foundation under
Grant No.~PHY89-04035, and
the Centre International des Etudiants et Stagiaires de France.

\end{document}